\documentclass[11pt]{article}

\usepackage{float}
\usepackage{placeins}
\usepackage{amsmath}
\usepackage{color}
\usepackage{amssymb}
\usepackage{mathtools}
\usepackage{subfigure}
\usepackage{multirow}
\usepackage{epsfig}
\usepackage{listings}
\usepackage{enumitem}
\usepackage{graphicx}
\usepackage{graphics}
\usepackage{epstopdf}
\usepackage{longtable}
\usepackage{hyperref}
\usepackage{breakurl}
\usepackage{epigraph}
\usepackage{xspace}
\usepackage{amsfonts}
\usepackage{eurosym}
\usepackage{ulem}
\usepackage{comment}
\usepackage{setspace}
\usepackage{geometry}
\usepackage{caption}
\usepackage{pdflscape}
\usepackage{array}
\usepackage[round]{natbib}
\usepackage{booktabs}
\usepackage{dcolumn}
\usepackage{url}
\usepackage{mdframed}
\usepackage{mathrsfs}

\usepackage{enumitem}

\usepackage{xcolor}
\hypersetup{
    colorlinks,
    linkcolor={blue!50!black},
    citecolor={blue!50!black},
    urlcolor={blue!50!black}
}
\definecolor{col3}{rgb}{0,.27,.4375} 

\newcolumntype{d}[1]{D{.}{.}{#1}} 

\newcolumntype{L}[1]{>{\raggedright\let\newline\\arraybackslash\hspace{0pt}}m{#1}}
\newcolumntype{C}[1]{>{\centering\let\newline\\arraybackslash\hspace{0pt}}m{#1}}
\newcolumntype{R}[1]{>{\raggedleft\let\newline\\arraybackslash\hspace{0pt}}m{#1}}

\newcommand{\Sref}[1]{Section~\ref{sec:#1}}

\newcommand{\Cref}[1]{Corollary~\ref{coro:#1}}
\newcommand{\cref}[1]{Cor.~\ref{coro:#1}}

\newcommand{\ie}{{\it i.e.}\ }
\newcommand{\eg}{{\it e.g.}\ }
\newcommand{\etc}{{\it etc.}\xspace}

\newcommand{\flabel}[1]{\label{fig:#1}}
\newcommand{\fref}[1]{Fig.~\ref{fig:#1}}
\newcommand{\Fref}[1]{Figure~\ref{fig:#1}}

\newcommand{\be}{\begin{equation}}
\newcommand{\ee}{\end{equation}}
\newcommand{\bea}{\begin{eqnarray}}
\newcommand{\eea}{\end{eqnarray}}
\newcommand{\beaa}{\begin{align}}
\newcommand{\eeaa}{\end{align}}

\newcommand{\bi}{\begin{itemize}}
\newcommand{\ei}{\end{itemize}}

\numberwithin{equation}{section}

\begin{document}

\begin{titlepage}
\title{The Distributional Short-Term Impact of the COVID-19 Crisis on Wages in the United States}
\author{Yonatan Berman\footnote{London Mathematical Laboratory, The Graduate Center and Stone Center on Socio-Economic Inequality, City University of New York,~\url{yberman@gc.cuny.edu}}\,\, \thanks{I wish to thank Janet Gornick and Salvatore Morelli for helpful comments.}}

\date{\today}

\maketitle

\begin{abstract}
\noindent 
This paper uses Bureau of Labor Statistics employment and wage data to study the distributional impact of the COVID-19 crisis on wages in the United States by mid-April. It answers whether wages of lower-wage workers decreased more than others', and to what extent. We find that the COVID-19 outbreak exacerbates existing inequalities. Workers at the bottom quintile in mid-March were three times more likely to be laid off by mid-April compared to higher-wage workers. Weekly wages of workers at the bottom quintile decreased by 6\% on average between mid-February and mid-March and by 26\% between mid-March and mid-April. The average decrease for higher quintiles was less than 1\% between mid-February and mid-March and about 10\% between mid-March and mid-April. We also find that workers aged 16--24 were hit much harder than older workers. Hispanic workers were also hurt more than other racial groups. Their wages decreased by 2--3 percentage points more than other workers' between mid-March and mid-April.
\\

\bigskip
\end{abstract}
\setcounter{page}{0}
\thispagestyle{empty}
\end{titlepage}
\pagebreak \newpage


\section{Introduction} \label{sec:introduction}

The COVID-19 outbreak led to an unprecedented abrupt shock to many countries, with the United States becoming the epicenter of the pandemic in late March 2020. With more than a million COVID-19 cases, and hundreds of millions of people in lockdown, the economic impact of the pandemic is massive and may become even greater in the near future.
 
Among the early outcomes of the ``Great Lockdown'' that followed the outbreak of the pandemic were the shutdown of many businesses and large-scale layoffs of workers. Millions of US workers have filed initial unemployment claims, bringing the unemployment rate in the US to 15.7\%~\citep{DOL2020}. The labor market impact of the lockdown is expected to be worst for jobs that cannot be performed remotely and are not considered essential. Such jobs are also generally lower-paid~\citep{AdamsBonevaGolinRauh2020}. Early evidence of the impact on unemployment suggests that it rises faster for women, Black men and Hispanics~\citep{AdamsBonevaGolinRauh2020,Kalev2020}.~\citet{mongey2020workers} find ``that workers in low-work-from-home or high-physical proximity jobs are more economically vulnerable.''

This paper studies the short-run distributional impact of the COVID-19 crisis on wages. In particular, it answers whether, indeed, wages of lower-wage workers decreased more than others' and to what extent. It also clarifies the magnitude of the effect -- what share of the labor force was substantially impacted? The paper also addresses age-, gender-, and race-specific impact. It answers whether wages of young workers decreased more than of older workers, whether wages of female workers were hurt more than of male workers, and how the impact on wages depended on race.

To answer these questions one would have to compare wages of employees just before the coronavirus pandemic outbreak in the US, to their wages at a later time. Yet, reliable large-scale real-time tracking of individual wages is challenging. Most administrative data cannot be used in real-time. Most large-scale surveys report earnings on a quarterly or annual basis. Several ad-hoc surveys have been conducted in the US and elsewhere~\citep{BartikETAL2020,AdamsBonevaGolinRauh2020}, yet their scale is limited and it is difficult to verify how representative their samples are for the labor market. Moreover, most of these data sources are not longitudinal, making it difficult to follow individuals over time. This paper uses wage and employment data from the monthly establishment survey of the Bureau of Labor Statistics (BLS, see~\citet{BLS2020}). These data represent 110--130 million private sector workers, divided into industry sectors.

Combining employment and wage data we create synthetic panels for private sector workers, from which we estimate non-anonymous growth incidence curves (NAGICs, see~\citet{bourguignon2011non}). NAGICs quantify the average wage growth rates of individuals occupying a specific wage rank at the beginning of a given period.
Thus, constructing a NAGIC for US workers for a period that spans from just before the outbreak of the pandemic in the US and a later point in time would quantify its impact on wages at different points along the wage distribution.

The NAGICs show that the weekly wages of workers at the bottom quintile decreased by 6\% on average between mid-February and mid-March and by 26\% between mid-March and mid-April. The rest, \ie top 80\%, lost less than 1\% on average by mid-March and about 10\% by mid-April. For comparison, between January and February, all quintiles, including the bottom, gained about 0.6\% of their wages on average.

We find no evidence for a significant change on hourly wages for workers that remained employed. Thus, the major reduction of weekly wages among workers at the bottom quintile is due to lay offs and reductions in working hours. It is not, at least by mid-April, due to changes in hourly wages.

Combining the panels with data from the Current Population Survey (CPS) on the composition of industry sectors by age, gender and race, we study the impact of the COVID-19 outbreak on the wages of different subgroups. We find that the reduction in weekly wages following the COVID-19 outbreak was slightly more pronounced among women than among men. Wages of female workers decreased by 0.2--0.3 percentage points more than of male workers. The decrease in wages was particularly large for younger workers, aged 16--24, in comparison to older workers.

The decrease in weekly wages was also large among Hispanic or Latino workers. The average reduction in weekly wages between mid-February 2020 and mid-March 2020 was greater by 0.4 percentage points among Hispanic workers than among all workers, and by 0.5 percentage points within the bottom quintile. Between mid-March and mid-April the reduction in wages among Hispanic workers was greater by about 2 percentage points than among all workers, and by 3 percentage points within the bottom quintile. All subgroups considered (young, female, male, White, Black, Asian, and Hispanic workers) suffered from an average reduction of 4.8\%--6\% in weekly wages between mid-February and mid-March, and an average reduction of 25\%--30\% in weekly wages between mid-March and mid-April within the bottom quintile.

The primary contribution of this paper is empirical. It provides new information on job losses and hourly and weekly wage losses among workers along the wage distribution following the COVID-19 outbreak. It also enables identifying the differential impact on wages by age, gender, and race. It joins the growing literature on labor market impact of the COVID-19 outbreak~\citep{alon2020impact,AdamsBonevaGolinRauh2020,BartikETAL2020,BartikETAL2020b,BrunoriETAL2020,CajnerETAL2020,chetty2020real,mongey2020workers}, focusing mainly on employment and the reaction of firms to the pandemic so far. It also joins the growing literature on how inequalities contribute to unequal health outcomes during the outbreak~\citep{ahmed2020inequality,glover2020health,van2020covid}.

From a methodological perspective the paper contributes to the labor and inequality literatures on short-run distributional changes, especially such that take into account panel data~\citep{kopczuk2010earnings,bourguignon2011non,piketty2018Distributional,berman2018growth}. Usually such research relies on large scale surveys, tax data or social security data, which cannot be used in real-time. This paper presents a way to use a reliable existing data source that is published on a monthly basis, the BLS establishment survey, to study short-run distributional changes.  This would allow to easily expand the analysis done here in the future. We also hope that this work will stimulate similar work in other countries with similar data sources.

The paper is organized as follows.~\Sref{gics} lays out necessary background on growth incidence curves, which serve as the primary tool for our analysis.~\Sref{data} describes the data we use.~\Sref{QNAGIC} presents the results on the distributional impact on wages, and~\Sref{genderrace} presents the impact on wages by age, gender, and race. We conclude in~\Sref{discussion}.

\section{Growth incidence curves} \label{sec:gics}

The main tool we will use to quantify and characterize the impact of the COVID-19 outbreak on wages is non-anonymous growth incidence curves (NAGICs). The concept of growth incidence curves (GICs) is central in studies of poverty and inequality. It is a useful tool for illustrating how income (or wealth, consumption, \etc) grows (or degrows) over a time period along the income (wealth, consumption, \etc) distribution. GICs quantify the average growth at each rank in the distribution. Yet, it matters how the ranks are defined and how incomes are compared, and there are two main types of GICs -- anonymous and non-anonymous.

Anonymous GICs~\citep{ravallion2003measuring} quantify the growth of the average income in the same income fractile over a time period. They ignore the identity of individuals within fractiles. In general, GICs are upward sloping, \ie lower for the bottom of the distribution and higher at the top, when inequality is increasing and vice versa. They are insensitive to mobility between income ranks.

Non-anonymous GICs~\citep{grimm2007removing,van2009income,bourguignon2011non} quantify the average income growth of individuals occupying a specific income rank at the beginning of a given period, but not necessarily the same income rank by the end of the period. They can be downward (upward) sloping depending on whether poorer individuals experience higher (lower) average growth rates than richer individuals. The NAGIC and the GIC thus differ through the former incorporating personal income mobility. They coincide when no reranking occurs between the initial and terminal dates.

While NAGICs are seemingly more informative of the individual experience of income changes, they require panel data, which are more difficult to collect, and are more sensitive to measurement error than anonymous GICs~\citep{Berman2020}. It is possible to overcome data limitations and create approximate NAGICs using quasi-NAGICs. In quasi-NAGICs individual incomes are not followed over time. Instead, the population is divided into groups and the average income within each group is followed over time. Mobility within each group is ignored, but not between groups. For example,~\citet{lakner2016global} apply this to the global income distribution in which the identity of a particular country-decile is preserved, ignoring changes in the identity of individuals within each country-decile.

We use a similar approach by dividing workers into industry sectors. This is done as follows: For two points in time, $A$ and $B$ (say two surveys, like February and March, or March and April), each sector $s = 1, 2, \ldots, S$ is characterized by the respective numbers of employed workers in the sector -- $n^A_s$, $n^B_s$ -- and their respective average wages (hourly or weekly) -- $w^A_s$, $w^B_s$. We create a panel of length $N = \sum_{i=1}^S \max\left\{n^A_i, n^B_i\right\}$ for the wages of the entire population of workers at times $A$ and $B$. The wages of each worker are set to either $w^X_s$ ($X$ being the point in time and $s$ being the industry sector of that worker) or 0, if the wages of $n^X_s$ workers were already set to $w^X_s$. This allows taking into account layoffs. If $n^B_s < n^A_s$, as expected in most sectors during a major crisis, then there will be workers whose wages at time $A$ were set to $w^A_s$, but at time $B$ to 0, which represent laid-off workers. This approximate panel now allows estimating NAGICs.

\section{Data} \label{sec:data}

Ideally, to quantify and characterize the distributional impact of the coronavirus pandemic on wages, one would have to compare wages of employees just before its outbreak, to their wages at a later time. Yet, large-scale real-time tracking of individual wages is practically impossible. Most administrative data sources cannot be used in real-time. Most large-scale surveys report earnings on a quarterly or annual basis. Several ad-hoc surveys exist, in the US~\citep{AdamsBonevaGolinRauh2020,BartikETAL2020} and elsewhere~\citep{AdamsBonevaGolinRauh2020}, but their scale is limited, and their focus is on employment rather than on wages.

We use wage and employment data from the establishment survey published monthly by the BLS~\citep{BLS2020}. Each month the BLS surveys US businesses on employment, hours, and earnings of employees. The data represent about 110--130 million workers, excluding about 30 million government workers, agricultural workers, self-employed workers whose businesses are unincorporated, unpaid family workers, and private household workers. The wage data do not include sources of income other than salaried work such as capital income or transfer payments, and they are also pre-tax.

The BLS data are not individual data, and are given by industry sector. Specifically, the available real-time (\ie monthly) data represent 14 broad sectors:\footnote{In some cases it is possible to create a finer picture based on more sectors, as discussed in Appendix~\ref{app:resolut}, without a substantial effect on the results.} Mining and logging, Construction, Durable goods, Nondurable goods, Wholesale trade, Retail trade, Transportation and warehousing, Utilities, Information, Financial activities, Professional and business services, Education and health services, Leisure and hospitality, and Other services. We use the number of employees in each sector, and the average hourly and weekly wages in each sector. In addition, we use the CPS data for 2019 to account for the share of employees in each sector by age, gender, and race~\citep{CPS2020}. These shares are assumed to have remained stable by April 2020.

For clarity and simplicity, we refer by the name of the month (\eg March 2020) to the data released for the specific month, although these data represent the pay period that includes the 12th of each month (usually the week that includes that day every month, see~\citet{BLS2020} Employment Situation Technical Note for more details). Thus, it is more accurate to treat the monthly data as representing the mid-point of each month. The February survey represents the situation just before the COVID-19 outbreak in the US. The March survey represents the initial impact of the evolving crisis. The April survey represents the major impact of the lockdown, imposed by most states in the US during the second half of March.

\section{Impact on wages along the wage distribution} \label{sec:QNAGIC}

\Fref{SumStat} presents the evolution of the total number of workers employed in the private sector, and their average weekly wages between January 2020 to April 2020. It shows only small differences between January, February and March. April, however, is substantially different. Almost 20 million jobs were lost between mid-March and mid-April. At the same time, the average wages of those still employed in April substantially increased -- from about \$980 per week to about \$1030 per week, an increase of 5\%. The black line in~\fref{SumStat} suggests that indeed, workers that were laid off in April were more likely to be low-paid, so that the average wages among workers still employed in April mechanically increased (see also Appendix~\ref{app:cps_gic}). If, instead, those who lost their job between March and April were included in the April workers population as earning zero wages, the average weekly wages in mid-April would be \$865 per week, a decrease of 11\% from mid-March.

\begin{figure}[!htb]
\centering
\includegraphics[width=1.0\textwidth] {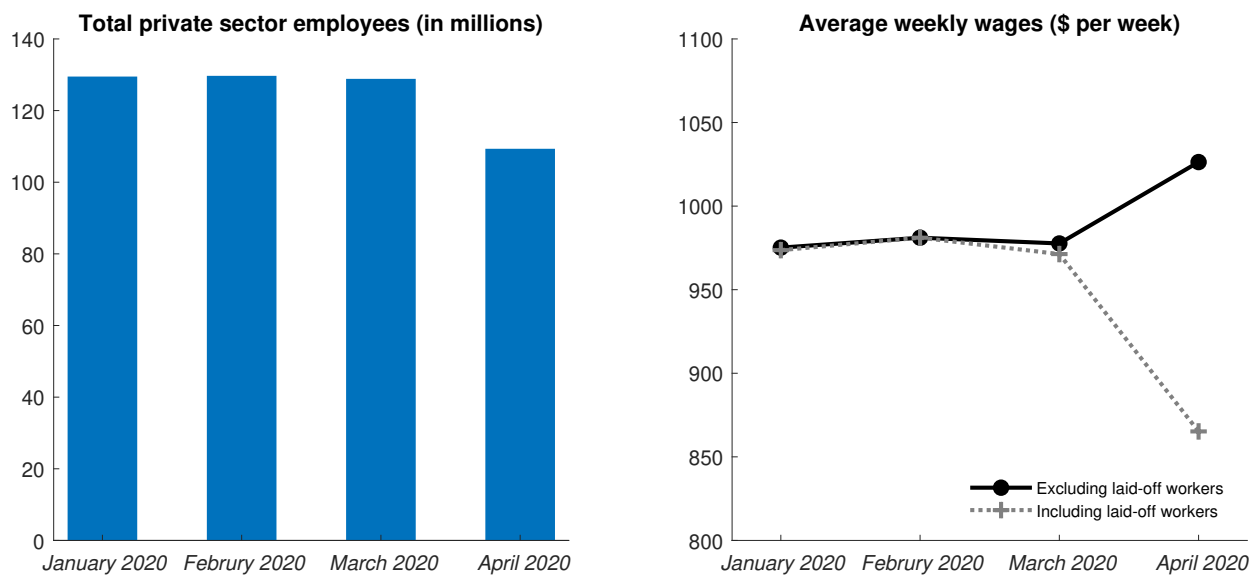}
\caption{Employment and average wages in the private sector, January 2020 -- April 2020. Left) Total number of employees on private nonfarm payrolls. Right) Average weekly earnings of all employees on private nonfarm payrolls. The black line includes only salaried workers in each month. The grey line includes laid-off workers as earning zero wages. Source:~\citet{BLS2020}.
\flabel{SumStat}}
\end{figure}

We aim to study a more refined distributional impact on wages using non-anonymous growth incidence curves. These will present the average change in wages along the distribution: for each quintile we estimate the average relative change in wages for the periods January--February, February--March, and March--April.\footnote{All the changes in wages we refer to are ``democratic'', \ie we consider the average relative change in wages within each wage rank, rather than the change in average wage within each wage rank~\citep{chenery1974redistribution,adamou2016,saez2019triumph}. The difference between the two can be substantial, in general. In this case, the difference between the two types of growth rate is small (see Appendix~\ref{app:nondemoc}).}

Specifically, since following individual wages is not possible, we construct quasi-NAGICs (see~\Sref{gics}). We use the number of employees and the weekly and hourly wages of employees in 14 different industry sectors.\footnote{For non-seasonally adjusted data it is possible to use a finer division to 32 sectors, and to 185 sectors until March. Yet, the differences between the resulting NAGICs with 14, 32, or 185 sectors are small (see Appendix~\ref{app:resolut}), and we would like to focus on the seasonally adjusted data.} By following the number of employees in each sector over time, layoffs are taken into account, as explained in~\Sref{gics}. We assume the same wage for all workers within each sector,\footnote{This assumption matters greatly to the wage distribution, of course, but not to the NAGICs. NAGICs only quantify the average growth of wages, and are therefore less sensitive to the shape of the distribution within each sub-group of the population. See Appendix~\ref{app:angics} for more details.} and that workers do not move between sectors during the time period in question. Specifically, we assume that laid-off employees were not rehired in another sector. These are assumptions imposed by the limitations of the available data. Yet, in practice, they only have a small impact on the estimated NAGICs (see Appendix).

\FloatBarrier

The procedure described results in a synthetic panel of all nonfarm private sector employees, each of which is characterized by her wages at two different points in time. The NAGICs are easily produced from these panels~\citep{bourguignon2011non}.~\fref{bls_nagic} presents the main results -- NAGICs for January--February, February--March, March--April. It shows that weekly wages of lower-wage workers were hurt more than of higher-wage workers between February and March, and most significantly between March and April. Lower-paid workers were more likely to see their wages and working hours decrease and to be laid off.

\begin{figure}[!htb]
\centering
\includegraphics[width=1.0\textwidth] {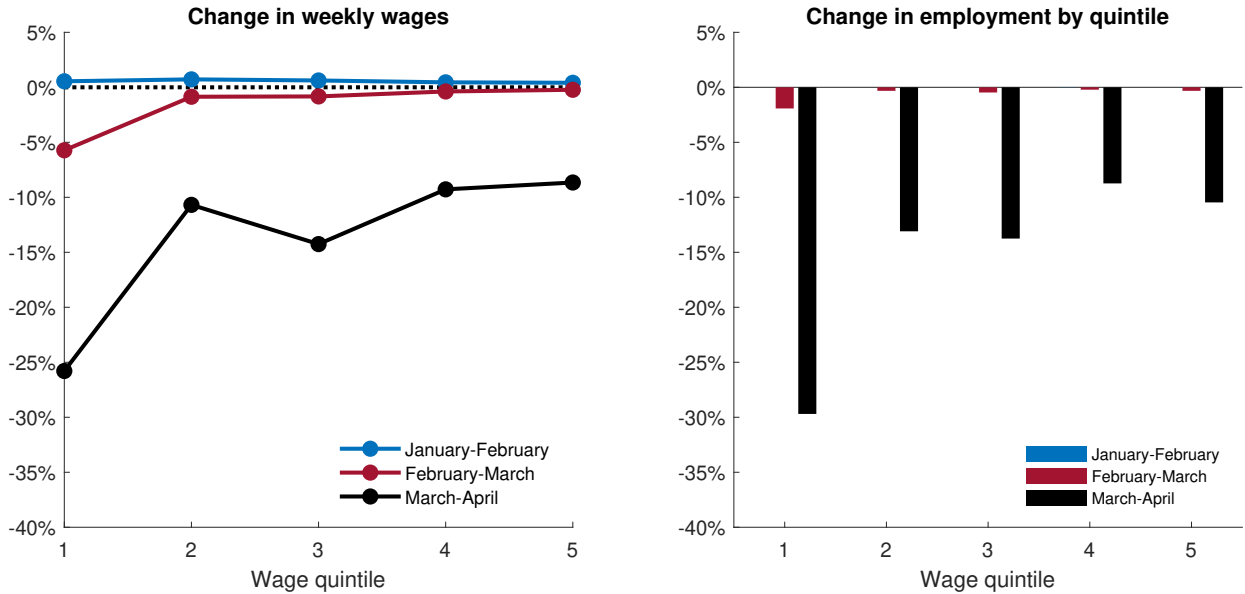}
\caption{Non-anonymous growth incidence curves for US private sector workers between January 2020 and April 2020 (left) and the change in employment by initial weekly wage quintile (right). The estimates are for three one-month periods: mid-January to mid-February (blue), mid-February to mid-March (red), and mid-March to mid-April (black).
\flabel{bls_nagic}}
\end{figure}

All wage ranks were negatively impacted between February and March, and between March and April. Between February and March the NAGIC was quite flat above the bottom quintile, and the decrease in wages was small for the top 80\%. Taking the January--February NAGIC as a reference, the COVID-19 impact is even more pronounced -- the typical monthly change in weekly wages tends to be small, and rather flat along the distribution. Not only that during the crisis the change is lower than this reference, it is significantly below 0, and especially low for workers at the bottom quintile.

The right panel in~\fref{bls_nagic} shows how changes in employment depended on the initial wage quintile for the three periods considered. There was no substantial change in employment in any quintile between January and February. Between February and March there is already a noticeable decrease, concentrated at the bottom quintile (inline with the findings of~\citet{BartikETAL2020}). The NAGICs in the left panel are clearly dominated by the changes in employment. Yet, weekly wages may also change for some due to increased or decreased number of working hours following the crisis, or the increase in demand for workers in specific sectors which may affect hourly wages.

The January--February and February--March NAGICs are almost insignificant compared to March--April. The average decrease in weekly wages in quintiles 2--5 was about 10\% by mid-April, and about 26\% in the bottom quintile. The total number of employees decreased between mid-March and mid-April by about 15\%. Yet, the reduction in employment was even more substantial for the bottom quintile. 30\% of the workers at the bottom quintile in mid-March were laid off by mid-April. Thus, workers at the bottom quintile were about three times more likely to be laid off compared to higher-wage workers.

\FloatBarrier

\section{Impact on wages by age, gender, and race} \label{sec:genderrace}

When producing the NAGICs, we created synthetic panels of employees. Employees were characterized by their wages (determined by their industry sector) at two different points in time, such as mid-February and mid-March. This panel can be augmented by combining it with data on the share of workers of different genders, age groups, and races in each sector~\citep{CPS2020}. This allows addressing questions of gender-, age-, and race-specific impact, \eg whether wages of female workers decreased more than males', or whether employees of a certain race were impacted more than of other races.

To explain how the panel is augmented, let us consider the case of female workers as an example. Suppose there are $n^A_s$ employees in sector $s$, at time $A$, of which a share $\rho_s$ are female employees, we randomly choose $\rho_s\cdot n^A_s$ employees, which are then flagged as corresponding to female employees in this sector. Repeating this for all sectors, we create a division of the population of employees to female and male, enabling the calculation of the average change in wages among female employees only and among male employees only. The random assignment means that each assignment will result in a slightly different average rate of change in wages. In each case we produced 1000 random assignments and consider the distribution of results to account for statistical significance. The uncertainty in these results is only due to the random assignment. It would not take into account potential biases which the data cannot account for, such as a preference to lay off women compared to men in some sectors, or vice versa. We also note that this method ignores the observation that within sectors there is a wage gap between men and women, White and non-White workers, \etc. Yet, the test presented in Appendix~\ref{app:angics} demonstrates that taking into account a distribution of wages rather than average wages does not have a large impact on the results.

This procedure was separately performed in each case, considering female and male workers, young workers (\ie workers aged 16--24) and then four different racial groups: White, Black or African American, Asian, and Hispanic or Latino. The results are presented in~\fref{barchanges}. It shows that between January 2020 and February 2020 all groups enjoyed an average increase in wages of about 0.6\%.

\begin{figure}[!htb]
\centering
\includegraphics[width=1.0\textwidth] {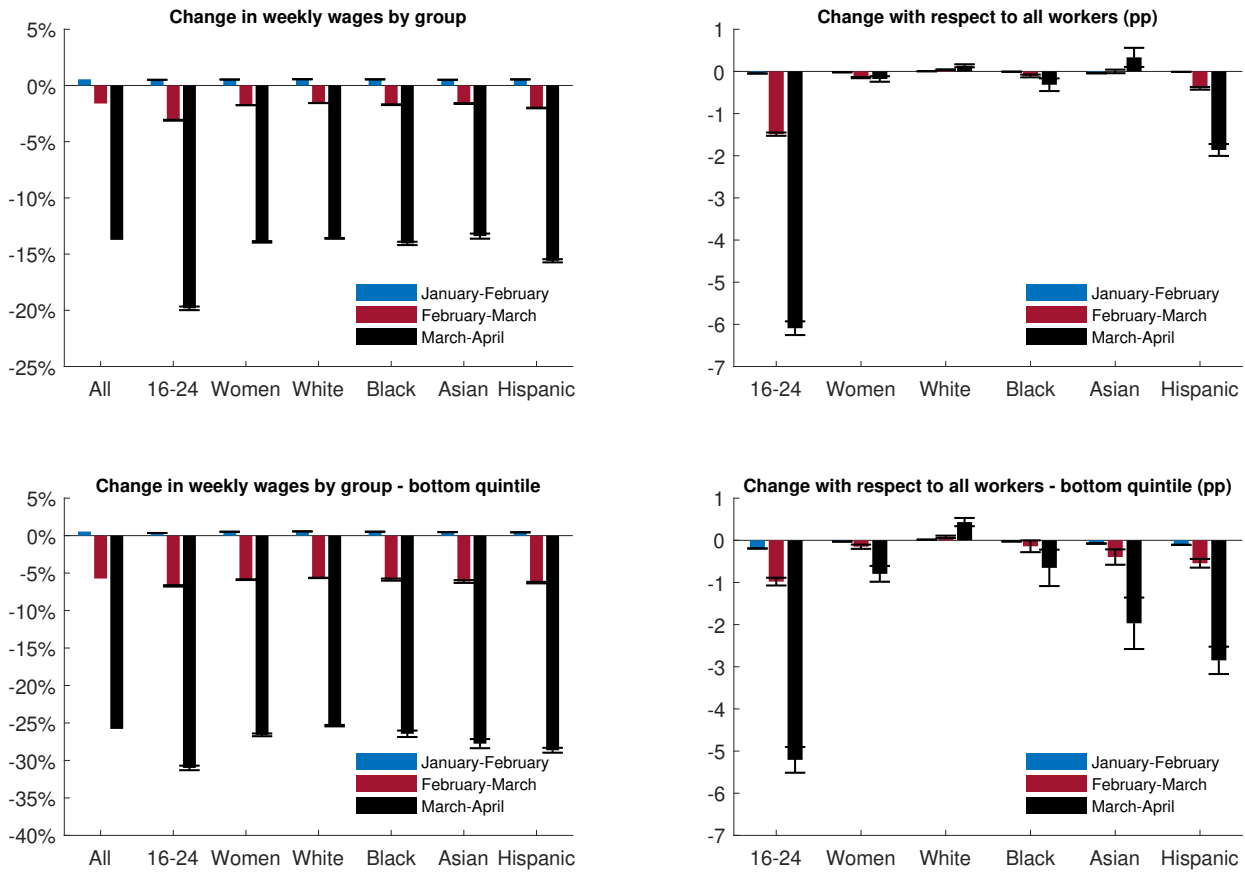}
\caption{Average relative change in weekly wages among different groups. Top left) The average relative change in wages in each group (all workers, young workers, women, Black, Asian, and Hispanic) for the periods January 2020 -- February 2020 (blue), February 2020 -- March 2020 (red), and March 2020 -- April 2020 (black); Top right) The difference between the average relative change in wages in each group and the change among all workers (in percentage points) for the periods January 2020 -- February 2020 (blue), February 2020 -- March 2020 (red), and March 2020 -- April 2020 (black); Bottom left and right) Similar to top but within the bottom quintile of the wage distribution. The error bars represent 95\% confidence bounds for the estimates produced through random assignments, as explained above.
\flabel{barchanges}}
\end{figure}

Between February 2020 and March 2020 the picture is strikingly different. First, all groups suffered from an average decrease in wages of about 1.4\%. However, women, Black, and particularly young workers and Hispanic workers suffered from significantly higher decreases. The decrease was slightly lower for White, although only mildly.

Focusing on workers at the bottom quintile in February, which, as~\fref{bls_nagic} shows, were hurt much more than others, the picture remains largely unchanged. All subgroups suffered from large decreases to their weekly wages. Again, Hispanic workers seem to have been hit more than other groups, with an average reduction of 5.5\% in their weekly wages. Weekly wages of young workers at the bottom quintile decreased by almost 6\%. Unlike the picture for the entire population of workers, Asian workers in the bottom quintile experienced relatively large decrease in wages. Hourly wages show very similar patterns (see Appendix~\ref{app:race_hou}).

The reduction in wages between mid-March and mid-April was clearly larger. All groups suffered an average reduction of 14\%--20\% in their weekly wages. As in the previous period, young workers and Hispanic workers are the subgroups that saw the biggest reductions in wages. Within the bottom quintile, all groups suffered an average reduction of 25\%--31\% in their wages. Asian and Hispanic workers, and particularly young workers suffered the largest declines.

The differences between racial groups reflect the composition of different industry sectors. For example, the leisure and hospitality sector, one of the sectors that suffered the biggest declines in employment between January and April, has a large share of Hispanic workers (24\%), but only a small share of Asian workers (6.9\%).

\FloatBarrier

\section{Conclusion}\label{sec:discussion}

The outbreak of COVID-19 in the United States had an unparalleled short-term impact on workers. Our results illustrate a major negative impact on wages by mid-April that was concentrated at, but not restricted to, the bottom quintile of workers. This impact is a combination of massive layoffs and a reduction in working hours, affecting mostly lower-paid workers.

Our analysis shows that the reduction in weekly wages following the COVID-19 outbreak was slightly more pronounced among women than among men. In particular, the decrease in weekly wages was pronounced among young workers (aged 16--24), relative to older workers, and to Hispanic workers, compared to other racial groups. These results highlight that while mortality rates from COVID-19 are very low among young adults, young workers have been economically hit harder than older workers due to the lockdown so far. This imposes additional political and social challenges to governments trying to manage the ongoing crisis.

The results call for continuous monitoring of the impact on employment and wages, which can inform policy trying to mitigate the negative impact of the lockdown on those hurt the most. Clearly, the US economy will gradually reopen during the next few months and many people will be rehired or employed in other businesses and sectors. Wages may start increasing. Yet, it is still unclear when this will happen, and how quick, if at all, the recovery will be, especially for those hit hardest during the past few months.

\clearpage

\bibliography{mobmob}

\clearpage

\appendix

%
%

\section{Sensitivity to resolution of industry sectors}
\label{app:resolut}

The NAGICs presented above rely on a division of workers into sectors. At this early stage the BLS seasonally adjusted establishment data for April are only available at a rather coarse resolution, and are based on 14 industry sectors: Mining and logging, Construction, Durable goods, Nondurable goods, Wholesale trade, Retail trade, Transportation and warehousing, Utilities, Information, Financial activities, Professional and business services, Education and health services, Leisure and hospitality, and Other services. The non-seasonally adjusted data are available at a slightly finer resolution, with some of the 14 sectors broken down into sub-sectors. The finer resolution allows dividing all workers into 32 sectors. For the periods January-- February and February--March it is possible to use an even finer resolution of 185 sectors.

Clearly, the finer the division into sectors, the more accurate the resulting NAGICs are, since we rely on the sector-average wage as being representative of the wages of workers in each sector. Since the seasonally adjusted data provide more relevant information for the interpretation of the NAGICs, we use the division to 14 sectors as the baseline NAGICs (see above). To demonstrate that this rather coarse division does not create a large impact on the resulting NAGICs, we compare the non-seasonally adjusted NAGICs based on a division to 14 sectors to a similar analysis in which workers are divided into 32 sectors and into 185 sectors. These results are presented in~\fref{bls_JanFebMar_democ_quint_14_185} and in~\fref{bls_NAGiC_32}. Indeed the differences between the NAGICs for the three resolutions (14, 32, and 185 sectors) are small and do not change qualitatively any of the main results discussed above.

\begin{figure}[!htb]
\centering
\includegraphics[width=1.0\textwidth] {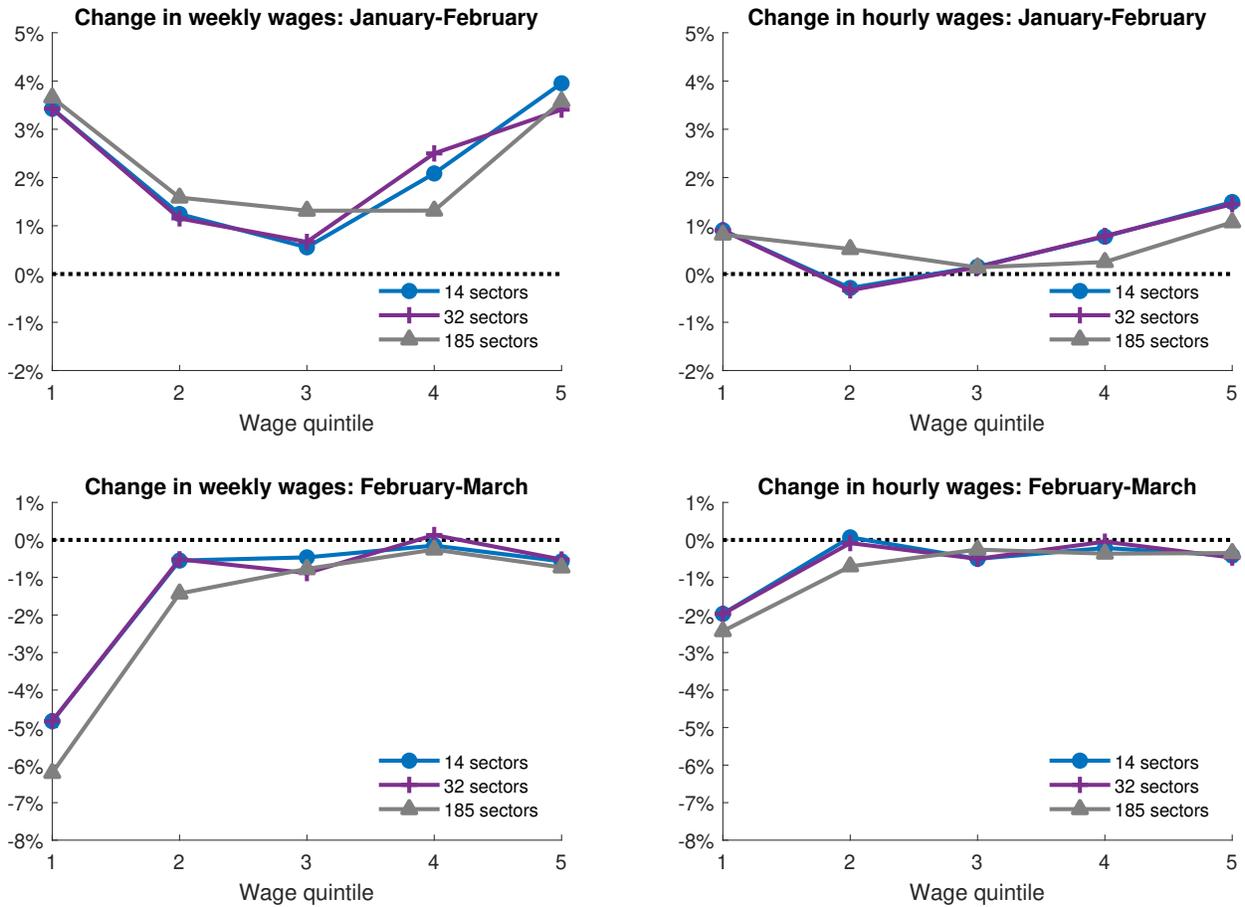}
\caption{Non-anonymous growth incidence curves for US private sector workers between January 2020 and February 2020 (top) and between February 2020 and March 2020 (bottom) by sector resolution, for weekly wages (left) and hourly wages (right). All curves are based on non-seasonally adjusted data.
\flabel{bls_JanFebMar_democ_quint_14_185}}
\end{figure}

\begin{figure}[!htb]
\centering
\includegraphics[width=1.0\textwidth] {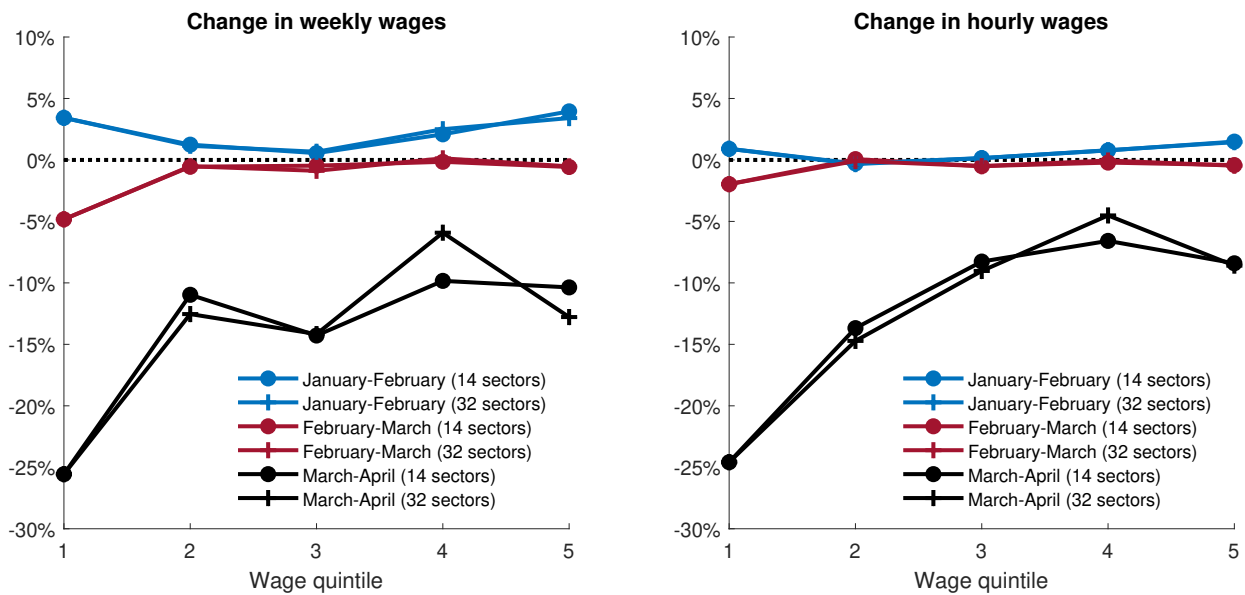}
\caption{Non-anonymous growth incidence curves for US private sector workers between January 2020 and April 2020. Left) For weekly ages; Right) For hourly wages. All panels are based on non-seasonally adjusted data.
The circles are based on a division to 14 sectors (Mining and logging, Construction, Durable goods, Nondurable goods, Wholesale trade, Retail trade, Transportation and warehousing, Utilities, Information, Financial activities, Professional and business services, Education and health services, Leisure and hospitality, and Other services).
The crosses are based on a division to 32 sectors (Mining and logging, Construction, Wood products, Nonmetallic mineral products, Primary metals, Fabricated metal products, Machinery, Computer and electronic products, Electrical equipment and appliances, Transportation equipment, Furniture and related products, Miscellaneous durable goods manufacturing, Food manufacturing, Textile mills, Textile product mills, Apparel, Paper and paper products, Printing and related support activities, Petroleum and coal products, Chemicals, Plastics and rubber products, Miscellaneous nondurable goods manufacturing, Wholesale trade, Retail trade, Transportation and warehousing, Utilities, Information, Financial activities, Professional and business services, Education and health services, Leisure and hospitality, Other services).
\flabel{bls_NAGiC_32}}
\end{figure}

\clearpage

\section{Changes in hourly wages by quintile}
\label{app:cps_gic}

Using the Current Population Survey (CPS)~\citep{CPSDAT2020} it is possible to find additional indications for the differential changes in hourly wages by quintile. The CPS does not follow individuals over time, and cannot be exactly matched with the information on layoffs at different sectors like the BLS data. Yet, it can be used to produce anonymous GICs for the periods January 2020 -- February 2020, February 2020 -- March 2020, and March 2020 -- April 2020 for hourly wages, while excluding workers that were laid off.

The results are presented in~\fref{cps_gic}. In January 2020 -- February 2020 and February 2020 -- March 2020 the GICs are nearly flat, indicating that the changes in hourly wages were rather homogeneous across the wage distribution. The March 2020 -- April 2020 the picture is different. First, average hourly wages increased substantially for all quintiles, apart from the top quintile. Second, the increase in hourly wages generally decreases with the wage rank. Similar to~\fref{SumStat}, this reaffirms that lower-paid workers were more likely to be laid off, which mechanically leads to an increase in hourly wages, even if workers' wages have not changed.

\begin{figure}[!htb]
\centering
\includegraphics[width=0.6\textwidth] {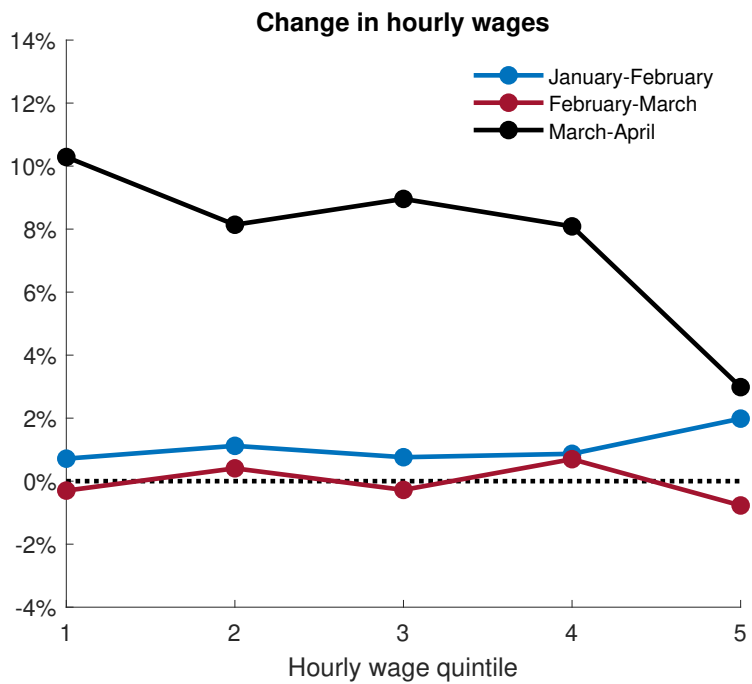}
\caption{Anonymous growth incidence curves for US workers between January 2020 and April 2020, based on CPS hourly wage data.
\flabel{cps_gic}}
\end{figure}

\clearpage

\section{Average change in wages and change of average wages}
\label{app:nondemoc}

The change in wages among different groups can be quantified in various ways. The most common way is by the relative change in average wages. This is similar to how income growth is typically measured. A major advantage of using this measure is that it does not require information on how wages are distributed within a certain group. Only averages are needed. Another common way to quantify the change in wages is by averaging over individual relative changes in wages. For each individual in a certain group a growth rate is estimated, and then the change in wages of that group is defined as the average across all individual growth rates. The advantage of this approach is that it gives the same weight to each individual in the group, rather than to each dollar earned by a member of the group, like the previous approach.

The differences between the two approaches have been thoroughly discussed in the literature~\citep{chenery1974redistribution,atkinson1975economics,adamou2016,saez2019triumph}. In the context of domestic product growth, the former approach is the standard way in which GDP growth is quantified. The latter is sometimes referred to as ``democratic'', or as ``People's growth''~\citep{saez2019triumph}. Since we aim to quantify as best as possible the experience of individual change in wages, the results in~\fref{bls_nagic} and~\fref{barchanges} are quantified using the average growth rate. The differences between the two approaches in the context of this analysis are very small. This is illustrated in~\fref{BLS_JAN_FEB_MAR_nondemoc}. It shows the NAGICs as in~\fref{bls_nagic}, also using the growth of average wages in each quintile rather than only the average growth rate.

\begin{figure}[!htb]
\centering
\includegraphics[width=1.0\textwidth] {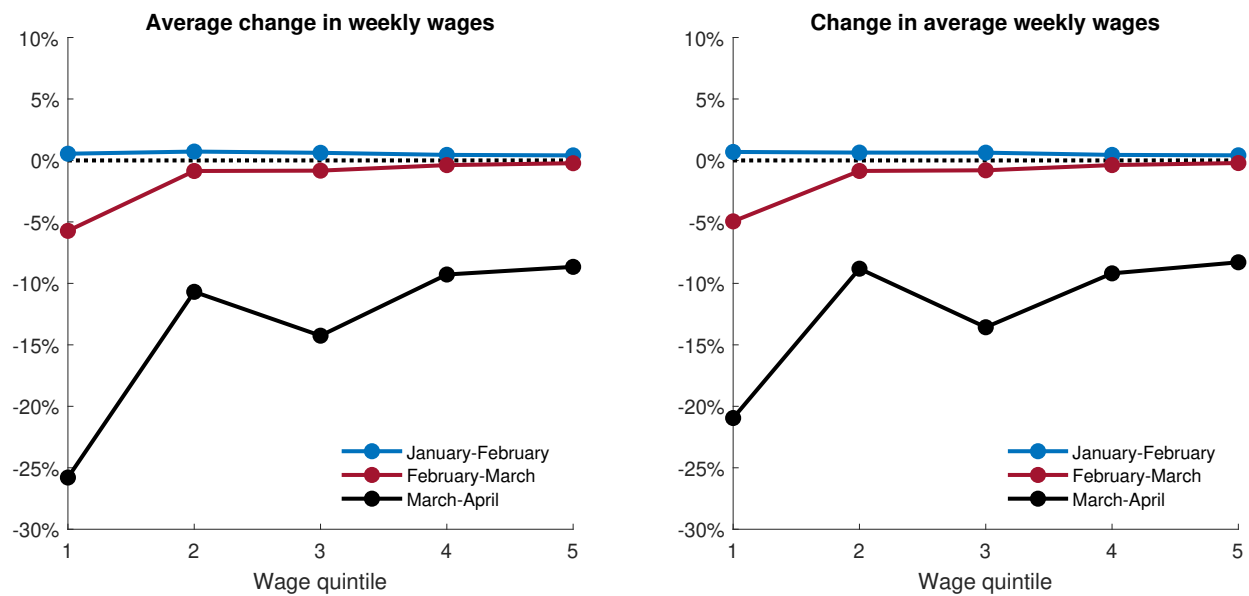}
\caption{Non-anonymous growth incidence curves for US private sector workers between January 2020 and April 2020. Left) Using the average relative change in weekly wages (same as~\fref{bls_nagic}); Right) Using the relative change in average wages in each quintile.
\flabel{BLS_JAN_FEB_MAR_nondemoc}}
\end{figure}

\clearpage

\section{The impact of using sector-average wages}
\label{app:angics}

The main limitation of the quasi-NAGIC approach is that the wage of each worker in the synthetic panel is taken to be the average wage in the sector to which she belongs. This is a limitation since wages differ dramatically within sectors. High-rank employees in some companies, whose wages are among the highest in the entire labor force, might be grouped together with employees that are earning the minimum wage. Nevertheless, since we are interested in average changes in wages, and not in the shape of the wage distribution, this limitation may not have a big impact on the estimated NAGICs, and the sector-average wages may still be indicative of the wages of most workers in the sector.

To test this we simulate a distribution of wages within each sector. Within each sector we assume the wage distribution is lognormal, a rather standard modeling assumption,\footnote{See~\citet{pinkovskiy2009parametric}. Also~\citet{benabou2000unequal} argued that ``the lognormal is a good approximation of empirical income distributions, leads to tractable results, and allows for an unambiguous definition of inequality.''} and assume the same degree of inequality within each sector. We simulate the distribution within each sector independently in each month, so we can no longer consider the synthetic panel -- the correlation between wages at the two months will no longer represent the realistic correlation, which is close to 1~\citep{kopczuk2010earnings}. Thus, we use anonymous GICs, in which the panel component does not play a role. We compare the GICs produced using sector-average wages and the GICs produced when simulating a realistic wage distribution. Since each randomly-generated simulation yields a slightly different GIC, we consider the average of 1000 simulations, noting that in all cases described below the variation between GICs in different realizations was small.

\Fref{BLS_JAN_FEB_MAR_anon_logn} presents the comparison considering weekly wages using the same data used for~\fref{bls_nagic}. It shows that for the period January--February the sector-average and the simulated anonymous GICs are indistinguishable. The differences for the period February--March are noticeable, but very small. For March--April the changes are slightly larger, but still smaller than 5 percentage points for all quintiles. The differences do not suggest any of the conclusions discussed above is invalid. We also note that the sensitivity of these results to the degree of inequality assumed in the simulated GICs is small within the realistic domain of Gini coefficients, between 0.3 and 0.5. While the simulated GICs may be more realistic than the sector-average based GICs, we remind the reader that our goal is to provide NAGICs, which are more informative on the individual experience of wage changes than anonymous GICs.

\begin{figure}[!htb]
\centering
\includegraphics[width=0.6\textwidth] {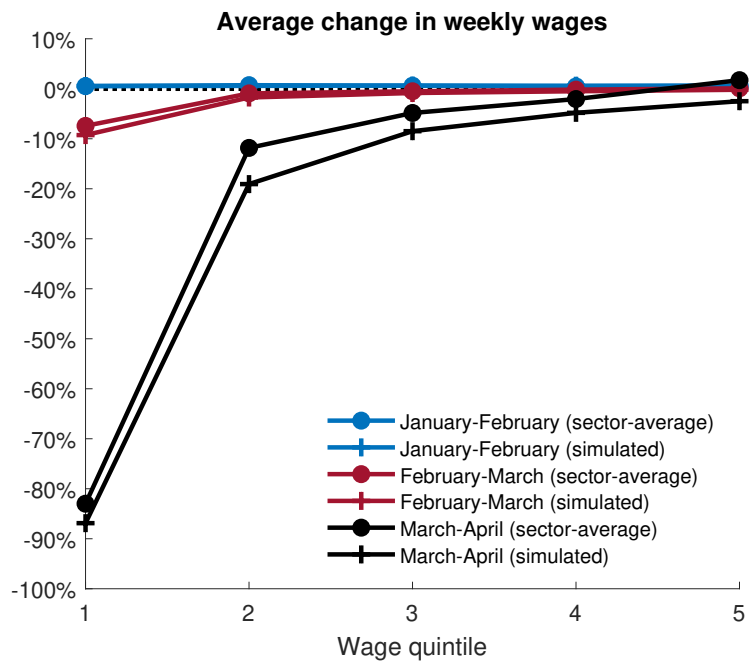}
\caption{Anonymous growth incidence curves for US private sector workers between January 2020 and April 2020. The circles refer to the sector-average GICs and the crosses to the simulated GICs (corresponding to a Gini coefficient of 0.4).
\flabel{BLS_JAN_FEB_MAR_anon_logn}}
\end{figure}

\clearpage

\section{Impact on hourly wages by age, gender, and race}
\label{app:race_hou}

\Fref{barchanges_hou} shows the impact on hourly wages by age, gender, and race, similarly to the results shown in~\fref{barchanges} for weekly wages. It shows a similar pattern. Between January 2020 and February 2020 all groups enjoyed a small average increase in wages. Between February 2020 and March 2020 all groups suffered from an average decrease in wages of about 0.5\%. The reduction in wages among women, Black, and particularly young workers and Hispanic workers was significantly higher than for other groups. The decrease was slightly lower for White workers. Focusing on workers at the bottom quintile in February, all subgroups suffered from large decreases to their hourly wages. Hispanic and Asian workers seem to have been hit more than other groups, with an average decline of 1.8\% in their hourly wages.

Between March 2020 and April 2020 all groups suffered a large decrease in wages, ranging from 12\% to 19\%. The reduction in wages was particularly significant among young workers and Hispanic workers. Within the bottom quintile all groups suffered reductions of 25\%--31\%. Young workers, as well as Asian and Hispanic workers were the groups hit hardest.

\begin{figure}[!htb]
\centering
\includegraphics[width=1.0\textwidth] {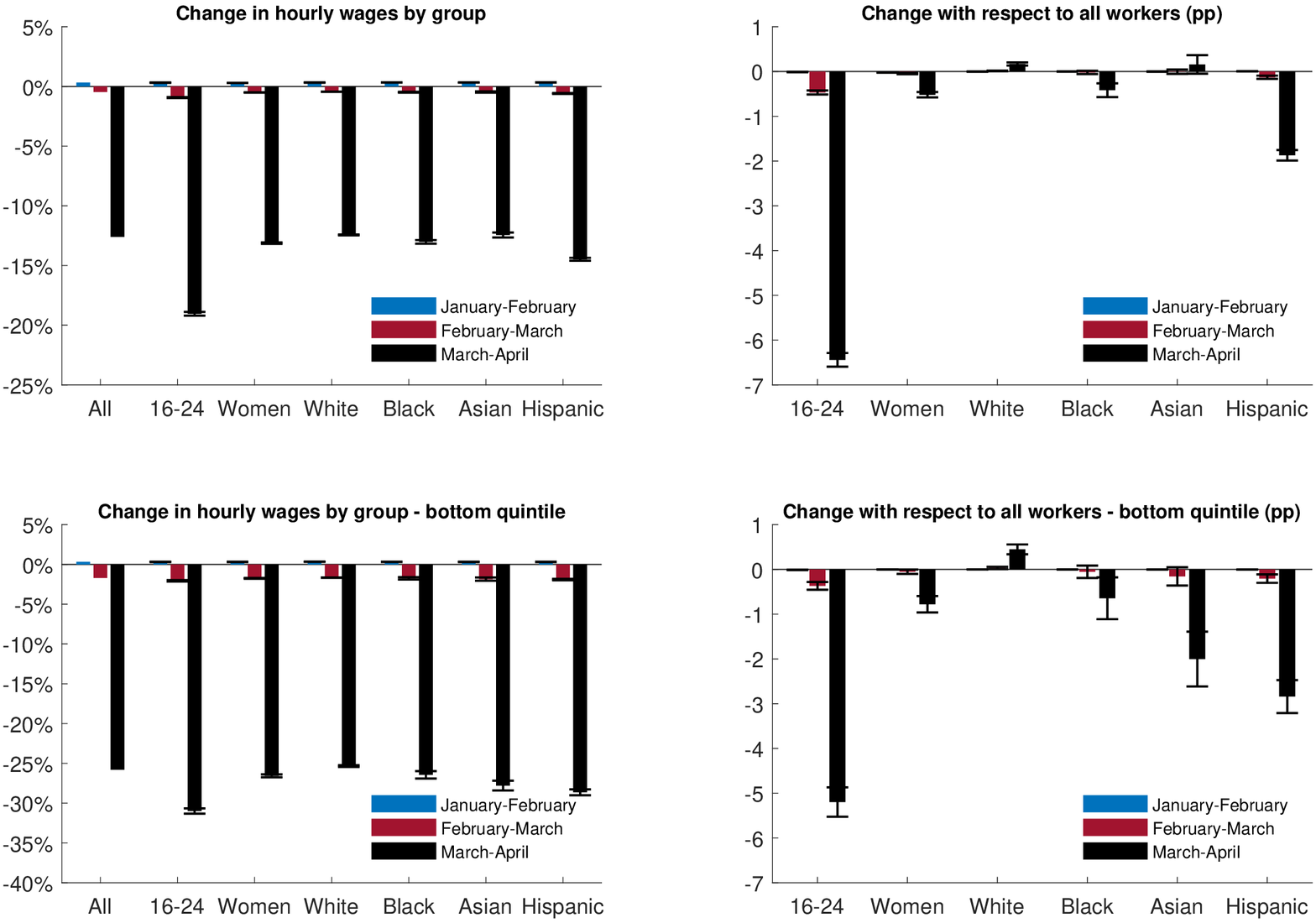}
\caption{Average relative change in hourly wages among different groups. Top left) The average relative change in wages in each group (all workers, young workers, women, Black, Asian, and Hispanic) for the periods January 2020 -- February 2020 (blue), February 2020 -- March 2020 (red), and March 2020 -- April 2020 (black); Top right) The difference between the average relative change in wages in each group and the change among all workers (in percentage points) for the periods January 2020 -- February 2020 (blue), February 2020 -- March 2020 (red), and March 2020 -- April 2020 (black); Bottom left and right) Similar to top but within the bottom quintile of the wage distribution. The error bars represent 95\% confidence bounds for the estimates produced through random assignments, as explained in~\Sref{genderrace}.
\flabel{barchanges_hou}}
\end{figure}

\end{document}